\documentclass[a4paper,fleqn,usenatbib]{mnras}

\usepackage[T1]{fontenc}
\usepackage{ae,aecompl}

\usepackage{graphicx}	
\usepackage{amsmath}	
\usepackage{amssymb}	

\def\submitted{AJ, accepted (arXiv:1812.00685)}

\title[Magellanic Clouds' young cluster metallicities]{Metallicity estimates of young
clusters in the Magellanic Clouds from Str\"omgren photometry of supergiant stars}

\author[A.E. Piatti et al.]{
Andr\'es E. Piatti$^{1,2}$\thanks{E-mail: andres@oac.unc.edu.ar}, Grzegorz Pietrzy\'nski$^3$, 
Weronika Narloch$^{4,5}$, Marek G\'orski$^4$ 
\newauthor and Dariusz Graczyk$^5$ \\
$^{1}$Consejo Nacional de Investigaciones Cient\'{\i}ficas y T\'ecnicas, Godoy Cruz 2290, C1425FQB, 
Buenos Aires, Argentina\\
$^{2}$Observatorio Astron\'omico de C\'ordoba, Laprida 854, 5000, 
C\'ordoba, Argentina\\
$^3$Nicolaus Copernicus Astronomical Center, 00-716 Warsaw, Poland\\
$^4$Departamento de Astronom\'{\i}a, Universidad de Concepci\'on, Casilla 160-C, Chile\\
$^5$Millennium Institute of Astrophysics, Santiago, Chile\\
$^6$Centrum Astronomiczne im. Miko$\l$aja Kopernika, PAN, Rabia\'nska 8, 87 - 100 Toru\'n, Poland
}

\date{Accepted XXX. Received YYY; in original form ZZZ}

\pubyear{2018}

\begin{document}
\label{firstpage}
\pagerange{\pageref{firstpage}--\pageref{lastpage}}
\maketitle

\begin{abstract}
We present results obtained from Str\"omgren photometry of 13 young ($\sim$30-220 Myr) Magellanic Cloud (MC) 
clusters, most of them lacking in the literature from direct metallicity measurements. We 
derived for them [Fe/H] values from a high-dispersion spectroscopy-based empirical calibration 
of the Str\"omgren metallicity sensitive index $m_{\rm 1}$ for yellow and red supergiants 
(SGs).  Particular care was given while estimating their respective uncertainties. In order to
obtain the mean cluster metallicities, we used [Fe/H] values of selected SGs for which we required 
to be located within the cluster radii, placed in the expected SG region in the cluster
colour-magnitude diagrams, and with [Fe/H] values within the FWHM of the observed cluster metallicity
distributions. The resulting metallicities for nearly 75 per cent of the cluster
sample agree well with the most frequently used values of the mean MCs' present-day metallicities.
The remaining clusters have mean [Fe/H] values that fall near the edge of the MC present-day
metallicity distributions.  When comparing the cluster metallicities with their present positions, 
we found evidence that supports the claimed recent interaction of 
the MCs with the Milky Way, that could have caused that some clusters were scattered from their birthplaces. Indeed, we show examples of clusters with metal contents typical of the galaxy inner
regions placed outward them. Likewise, we found young clusters, at present located
in the inner regions of both MCs, formed out of gas that has remained unmixed since several
Gyr ago.
\end{abstract} 




\section{Introduction}

Nearly 100-200 Myr ago the Milky Way has experienced its first passage to the Magellanic Clouds 
(MCs) \citep{beslaetal2012}. As a consequence of such an interaction sudden cluster
formation episodes have taken place throughout these galaxies \citep{bch05,mk2011,p18c}.
Since clusters share the metallicities of their birthplaces, those younger objects can tell us
about the efficiency of the gas mixing within the MCs, the metal enrichment due to
the galaxy chemical evolution, the infall of gas from MCs-Milky Way interaction, etc. Young clusters
also describe the most recent structures of these galaxies, where active cluster formation regions 
can even exist. Young clusters are tracers of the galaxy present-day metallicity distributions. 
By analysing the broadness of such a metallicity distributions and their relationship with the young
cluster spatial distribution, we can get some clues about the effectiveness of scattering clusters
from galaxy interactions, to assess whether clusters have been formed in an outside-in or inside-out
formation scenario, among others.

The number of young clusters (age $\la$ 200 Myr) in the MCs with actual measurements of their
metal contents is really negligible in the literature. Most of the catalogued young clusters have 
been studied photometrically, using their colour-magnitude diagrams (CMDs) to derive their ages
by assuming that they share the known MCs' mean present-day metallicities 
\citep[see, e.g.][]{getal10,p17e}. Sometimes, a couple of different [Fe/H] values have been chosen
to match theoretical isochrones to the cluster CMDs. More recently, bayesian and 
maximum-likelihood approaches have been implemented to fits thousand of isochrones to the CMDs in order
to get the best fitted cluster ages and metallicities \citep{detal14,pvp15}. Nevertheless, none of
them perform direct measures of the cluster's members chemical compositions.

With the aim of mitigating the lack of metallicity measurements of young MC clusters, we
used here  Str\"omgren photometry of yellow and red supergiants (SGs) to provide for the
first time with accurate mean [Fe/H] values for 12 young MCs, and for the Small Magellanic Cloud cluster
NGC\,330, whose previous spectroscopic iron abundance served as a reference for our
metallicity scale. Details of the data sets obtained and the careful process
carried out with the images until obtaining the standardised Str\"omgren photometry is described in 
Section 2. In Section 3 we deal with the cluster metallicities, how we derived them and
thoroughly estimated their uncertainties. We analyse and discuss in Section 4 different implications 
of the resulting cluster [Fe/H] values, in the context of the MCs' chemical evolution histories
and interaction with each other and of them with the Milky Way. Finally, Section 5 summarises
the main conclusions of this work.

\section{Str\"omgren photometry data set }

The photometric data sets analysed in this work were obtained during an observing campaign
aimed at studying the chemical evolution of the MCs from star clusters and field stars
(programme ID: SO2008B-0917, PI: Pietrzy\'nski). The images are publicly available at the National Optical Astronomy Observatory (NOAO) 
Science Data Management (SDM) Archives.\footnote{http //www.noao.edu/sdm/archives.php.} Two different observing runs were
carried out (17-19 December 2008 and 16-18 January 2009) with the SOAR Optical Imager (SOI)
attached to the 4.1m Southern Astrophysical Research (SOAR) telescope
(FOV = 5.25$\arcmin$$\times$5.25$\arcmin$, scale=0.154$\arcsec$/px in binned mode). The images resulted
of excellent quality (typical FWHM $\sim$ 0.6$\arcsec$) and were processed following
the SOI's pipeline guidance available at http://www.ctio.noao.edu/soar/content/soar-optical-imager-soi. In doing this, we used suitable zero and flat-field images obtained during each
observing night. Table~\ref{tab:table1} lists the log of observations for the
studied young MC clusters. Other subsample of clusters have been analysed previously to
search for intrinsic metallicity spreads among Large Magellanic Cloud (LMC) old globular
clusters \citep{pk2018} and in NGC\,1978  \citep{martocchiaetal2018b,pb2018} and
hints of multiple populations among Small Magellanic
Cloud (SMC) intermediate-age clusters \citep{niederhoferetal2017,p18b}.

We selected the standard stars HD64, HD3417, HD12756, HD22610, HD57568, HD58489, HD66020,
TYC 7547-711-1, TYC 7548-698-1, TYC 7583-1011-1, TYC 7583-1622-1, TYC 7626-763-1,
TYC 8033-906-1, TYC 8067-207-1, TYC 8104-856-1 and TYC 8104-969-1 \citep{hm1998,p2005}
to secure transformation of the instrumental magnitudes to the standard system.
Particular care was given to the observations of these stars by obtaining images
in all the $uby$ filters at small and large hour angle (airmass between 1.02 and 2.20).
Additionally, we observed each star twice at a given airmass, with the aim of placing them
in each of the two CCDs used by SOI. As shown in \citet{pb2018}, there is an excellent agreement
between the independent transformation coefficients from both CCDs. For this reason, we decided 
to use all the measured stars, regardless their positions in SOI. The transformation equations
 fitted are as follows:

$v = v_1 + V_{\rm std} + v_2\times X_v + v_3\times (b-y)_{\rm std} + v_4\times m_{\rm 1 std}$,\\

$b = b_1 + V_{\rm std} + b_2\times X_b + b_3\times (b-y)_{\rm std}$,\\

$y = y_1 + V_{\rm std}  + y_2\times X_y + y_3\times (b-y)_{\rm std}$,\\

\noindent where  $v_i$, $b_i$ and $y_i$ are the i-th fitted
coefficients, and $X$ represents the effective airmass. The resulting coefficients
are listed in Table~\ref{tab:table2}.

The instrumental magnitudes were derived from point-spread-function (PSF) photometry
using the routine packages {\sc daophot}, {\sc allstar}, {\sc daomatch} and {\sc daomaster}
in their stand-alone version \citep{setal90}. The PSF of each image was created
from a sample of nearly one hundred not-saturated, bright,
isolated stars, interactively selected and distributed throughout the entire image.
These PSF samples were previously cleaned from fainter neighbours using preliminary
PSFs built with the best nearly forty PSF candidates. We adopted a quadratically
spatially-varying PSF function for all the images. We applied the created PSFs
to the identified stellar sources and took advantage of the subtracted images
for identifying new fainter stars that were added to the previous list. The last
steps were iterated three times, deriving instrumental magnitudes from simultaneously
applying the respective PSF to the enlarged sample of stars. We computed
aperture corrections in the range -0.04 - -0.07 mag. Finally, we inverted
the fitted transformation equations to obtain magnitudes in the standard
system. 

Errors were estimated from extensive artificial star tests as previously
performed for other subsets of MC clusters imaged during the same
observing programme \citep[see][]{pk2018,p18b,pb2018}. In brief,
we used the stand-alone {\sc addstar} program 
in the {\sc daophot} package \citep{setal90} to add synthetic stars, 
generated bearing in mind the colour and magnitude distributions 
of the stars in the CMD  as well as the cluster radial stellar 
density profile. We added a number
of stars equivalent to $\sim$ 5$\%$ of the measured stars in order to avoid in the synthetic images 
significantly 
more crowding than in the original images. We created a thousand different images for 
each original one. We used the option of entering the number of
 photons
per ADU in order to properly add the Poisson noise to the star images. 
We then repeated the same steps to obtain the photometry of the synthetic images as described above, 
i.e., 
performing three passes with the {\sc daophot/allstar} routines. The photometric errors were derived from the magnitude difference between the output and input data
of the added synthetic stars using the {\sc daomatch} and {\sc daomaster} tasks. We found that this difference
resulted typically equal to zero and in all the cases smaller than 0.003 mag. The respective rms errors were 
adopted as the photometric errors.

\section{Str\"omgren metallicities}

\citet{gr1992} recommended the following expression to estimate metallicities
of SGs:

\begin{equation}
{\rm [Fe/H]} = \frac{(m_{\rm 1})_o + a_1 \times (b-y)_o + a_2}{a_3 \times (b-y)_o + a_4}
\end{equation}

\noindent where $a_1$ = -1.240$\pm$0.006, $a_2$ = 0.294$\pm$0.030, $a_3$ = 0.472$\pm$0.040
and $a_4$ = -0.118$\pm$0.020, respectively. Notice that $m_{\rm 1}$ = ($v-b$) - ($b-y$).
We used eq. (1) for cluster SGs that satisfy the following requirements: i) the SGs 
lie within the cluster radius \citep{betal08}. ii) They have intrinsic $(b-y)_o$ colours in 
the range 0.4 - 1.1 mag for which eq. (1) is valid. iii) They fall
above the cluster main sequence turnoff in the $V$ versus $b-y$ CMD, where cluster SGs are expected to be distributed. iv) Their individual [Fe/H]
values are within the FWHM of the metallicity distribution of all SGs complying with
the above criteria. This latter requisite helped us to clean the sample of cluster
SGs. Notice that field SGs are not homogeneously distributed throughout the observed
fields, so that the frequently procedure of choosing a region with an equal cluster area
far away from the cluster as a star field reference to clean the cluster CMD, could
be misleading. In addition, field SGs are distributed stochastically in the cluster CMD, 
so that it could not be straightforward to distinguish them from cluster SGs by  
considering only their positions in those CMDs. Fig.~\ref{fig:fig1} shows the CMDs
for all the stars within the clusters' radii with black dots, while selected stars above
the cluster turnoffs and with metallicities within the FWHM of the metallicity distributions 
are drawn with big black and red filled circles, respectively.

We extracted from the {\it Gaia} archive\footnote{http://gea.esac.esa.int/archive/} 
parallaxes ($\varpi$) and proper motions in Right Ascension (pmra) and Declination (pmdec) 
for stars located within 10 arcmin from the centres of our cluster sample, with the aim
of including an additional criterion on the membership status of cluster SG selection.
To choose cluster stars we constrained our sample to those satisfying the following criteria: 
i) stars located at the MC distances, i.e. $|\varpi|$ $<$ 3$\sigma(\varpi)$ and $|\varpi|$ $<$
4.0 mas. We rejected all stars with $\varpi$ not consistent with zero at more than 3$\sigma$
level \citep[see][]{vasiliev2018}; ii) stars located within the cluster radii \citep{betal08}. 
Unfortunately, we did not find stars with proper motion errors $\le$ 0.3 mas/yr, which correspond 
to $\sim$ 70 and 85 km/s, if the mean Large and Small Magellanic Clouds (L/SMC) distances are
used. Therefore, without the necessary proper motion accuracy, it was not possible to conduct
any membership probability analysis.

We have highlighted the sample of selected cluster SGs with big red filled circles in the 
cluster CMDs of Fig.~\ref{fig:fig1}. We also show their placement in the $(m_{\rm 1})_o$
versus $(b-y)_o$ plane, which includes iso-abundance lines according to eq. (1).
In order to estimate the individual metallicities, we first dereddened the measured
$b-y$ and $m_{\rm 1}$ colour indices by using the expression given by \citet{cm1976}
and the largest $E(B-V)$ value of those retrieved from the \citet[][hereafter H11]{hetal11} 
MC extinction map and from the NASA/IPAC Extragalactic Database (NED). For the sake of
the reader, Table~\ref{tab:table3} lists both $E(B-V)$ colour excesses. The uncertainties
in the [Fe/H] values were calculated by propagating every involved error,
namely: the photometric errors $\sigma (b-y)_o$ and $\sigma (m_{\rm 1})_o$ and the errors
in the $a_i$ values ($i= 1,..,4$) of eq. (1), according to the expression:\\

$\sigma{\rm [Fe/H]} =  [(\frac{(b-y)_o}{c}\sigma(a_1))^2 + (\frac{1}{c}\sigma(a_2))^2 +\\ \\(\frac{(b-y)_o{\rm [Fe/H]}}{c}\sigma(a_3))^2 + (\frac{{\rm [Fe/H]}}{c}\sigma(a_4))^2 + \\ \\
(\frac{(a_1 - a_3{\rm [Fe/H]})}{c}\sigma((b-y)_o))^2 + (\frac{1}{c}\sigma((m_{\rm 1})_o))^2]^\frac{1}{2}$,\\

\noindent where $c = a_3(b-y)_o +  a_4$. Since $\sigma{\rm [Fe/H]}$ varies from one SG to another within a cluster, 
we used the well-known maximum likelihood approach described in, e.g., \citet{pm1993} and
\citet{walker2006} to derive the mean cluster metallicities and the respective errors. The resulting [Fe/H] values
are listed in the last column of Table~\ref{tab:table3}.

\section{Analysis and discussion}

As far as we are aware, most of the studied clusters do not have direct estimates of their
metallicities. From a careful search through the available literature, we realised
that only NGC\,330 have been targeted for a spectroscopic metallicity analysis (see 
\citep[][and references therein]{gr1992}. \citet{gr1992}
obtained a mean value of [Fe/H] = -1.26 dex, in excellent agreement with our present 
estimate. Note, however, that the recent work by \citet{miloneetal2018} adopted a more metal-rich
value ([Fe/H] = -0.9 dex). \citet{dirschetal2000} estimated [Fe/H] = -0.57 dex for NGC\,1711,
rather different to our derived value (-0.06$\pm$0.05 dex). Notice that they showed a comparison of their metallicities
with those from high-dispersion spectroscopy that resulted in differences between 0.0 and 0.8 dex,
being their values more metal-poor. 
For the remaining clusters, previous photometric studies have
adopted the accepted mean galaxy present-day metallicities, i.e.. [Fe/H]= -0.4 and -0.7 dex, for
the LMC and SMC, respectively \citep[see, e.g.][]{pg13}. Some
few photometric studies have tried with a couple of different metallicity values
while matching isochrones to the cluster CMDs or recovering their formation histories (NGC\,376, 1844, 1847
and 2136). The middle columns of Table~\ref{tab:table3} list the values of ages and 
metallicities we found while searching the literature.

By comparing our resulting cluster metallicities with those previously used in the literature, 
we found some differences ($\Delta$([Fe/H]) $\sim$ 0.3-0.4 dex) that led us to speculate 
about the possibility that the assumption for young MC clusters to have metal contents similar 
to the mean galaxy present-day metallicities is justified for statistical purposes. Otherwise, 
when young clusters are studied to search for chemical abundance anomalies, light element
abundance variations, binary fraction, extended main sequence turnoffs, among others, the 
knowledge of their actual metallicities could have an impact. This could be the case, for
instance, of NGC\,1844, NGC\,1847 and NGC\,330, for which \citet{miloneetal13}, 
\citet{niederhoferetal15a} and \citet{miloneetal2018} adopted, respectively, more metal-rich 
metal abundances to show evidence of multiple populations. In order to see whether a better tracking of the split main sequences can be achieved, it would be worth trying to match
their CMDs with theoretical isochrones with metallcities similar to those derived in this work.
In the case of NGC\,1850, a cluster with a large population of near-critically rotating stars, 
a slightly more metal-rich value has usually been adopted \citep{bastianetal2017}.

The existence of a spread in metallicity within the younger stellar populations of both MCs is 
well-known. \citet{pg13}, using an homogeneous age/metallicity compilation, showed that the 
FWHM of such a scatter is 0.51 dex for the LMC and and 0.32 dex for the SMC 
\citep[see, also][]{chetal16,choudhuryetal2018}; the MC cluster populations also exhibit 
a noticeable scatter at their younger end \citep[see, also][]{perrenetal2017}. In this
context, most of the present studied clusters are within the expected metallicity range,
while others fall at the edge of the known metallicity distributions. This is the case of 
NGC\,330, the most metal-poor young SMC cluster ([Fe/H] = -1.15 dex) so far. In the LMC, 
NGC\,1847 resulted to be the most-metal poor young cluster ([Fe/H]=-0.91 dex) ever known, 
while NGC\,1711 turned out to be at the metal-rich end of the LMC cluster's metallicities
([Fe/H]= -0.06 dex). These relatively extreme metallicity values tell us that the
gas out of which these young clusters were formed was not well-mixed.  
Note that  there has been speculations about the role of
infall of unenriched (or less enriched) gas into the MCs leading to an unexpectedly
large spread in cluster abundances at a relatively constant age \citep[see, e.g.][]{dh98}.

We looked at the cluster positions in their respective host galaxies in order to search for 
any link of their derived metallicities with the chemical evolution histories of the MCs, particularly 
of those with more extreme values. Fig.~\ref{fig:fig2} depicts
with black points the spatial distributions in both MCs of all the clusters catalogued by 
\citet{betal08}. The studied clusters are drawn with big filled circles. As a spatial
reference, we have also included the areas defined by \citet{hz09} in the LMC main body 
(the bar is traced with a light-blue line) and the ellipses proposes by \citet{petal07d} 
as a simple representation of the orientation, dimension and shape of the SMC main body.
As can be seen, most of the studied LMC clusters are located along the bar, and some few 
others in the disc, while the studied SMC clusters are confined to the ellipse with
semi-major axis of $\sim$ 1 degree.

In an outside-in galaxy formation scenario -- which appears to be the case of both MCs 
\citep{meschin14,rubeleetal2018,p18c,p18d} -- the inner regions of a galaxy turn our to be 
more metal-rich than the outer ones. Indeed, from \citet{pg13} we found that the metallicity
level of field stars in the outer LMC disc  ($\rho >4 \degr$, $<$[Fe/H]$>$ = -0.90$\pm$0.20 dex)
is on average more metal-poor than that for inner disc field stars ($\rho <4 \degr$, 
$<$[Fe/H]$>$ =-0.50$\pm$0.20 dex). For the SMC, we got $<$[Fe/H]$>$ = -1.20$\pm$0.20 dex and
-0.70$\pm$0.15 dex for regions with semi-major axes larger and smaller than $1 \degr$, respectively \citep{p12a}. Star clusters share the metallicities of their birthplaces.
Nevertheless, with time they drift away from their birth locations. Interactions and other
perturbations may produce additional velocity components.

The derived metallicities of NGC\,330 ([Fe/H]= -1.15 dex, light-green circle in 
Fig.~\ref{fig:fig2}) and NGC\,1847 ([Fe/H]= -0.91 dex, orange circle in Fig.~\ref{fig:fig2})
are typical of stellar populations located in the outer regions of the MCs, although both
clusters are projected toward inner regions. Conversely, NGC\,1711 (red circle in 
Fig.~\ref{fig:fig2}), which is projected on to the LMC outer disc, have a metal content 
([Fe/H]= -0.06 dex) typical of the LMC bar. Recently, \citet{piattietal2018a} showed that 
a recently discovered young cluster placed in the outer disc of the LMC, possibly reached 
the present position after being scattered from the innermost LMC regions where it might have been born. This possibility could be applied to NGC\,1711, unless the cluster is the
outcome of an episode of recent cluster formation as a consequence of the first passage of 
the LMC by the Milky Way triggering cluster formation due to the ram pressure of
Milky Way halo gas \citep{piattietal2018b}. As for the birthplaces of NGC\,330 and NGC\,1847,
we can only infer that they have been formed from gas that has remained unmixed during the
last $\sim$ 4 Gyr in the SMC and $\sim$ 9 Gyr in the LMC. In order to infer these ages, we 
used the age-metallicity relationships derived by \citet[][see their figure 6]{pg13}; we then entered 
with the cluster metallicities and looked for the corresponding ages. 

Finally, we searched the literature for radial velocity (RV) measurements. As far as we are aware,
we found RVs for NGC\,330 \citep[149.0$\pm$8.0 km/sec][]{fb1980}, NGC\,1850
\citep[251.4$\pm$2.0 km/sec][]{fischeretal1993} and NGC\,2136  
\citep[271.4$\pm$0.4 km/sec][]{mucciarellietal2012}. 
Radial velocities are not available for the two anomalous LMC clusters NGC\,1711 and NGC\,1847.
One of the diagnostic diagrams most frequently used to assess whether a cluster belongs
to the LMC disc is that which shows the relationship between position angles (PAs) 
and RVs \citep{s92,getal06,shetal10,vdmareletal2002,vdmk14} for a disc-like rotation
geometry. We here followed the recipe used by \citet{s92}, who converted the 
observed heliocentric cluster RVs to Galactocentric RVs through eq.(4) in \citet{fw79}. 
We computed cluster PAs by adopting the LMC disc central coordinates 
obtained by \citet{vdmk14} from $HST$ average proper motion 
measurements for stars in 22 fields. We obtained PAs of 308$\degr$.0 and 44$\degr$.0
and Galactocentric RVs of 33.0 km/sec and 95.0 km/sec for NGC\,330 and 2136, respectively. 
These values are fully consistent with both clusters belonging to the LMC disc 
(see figure  7 in \citet{piattietal2018a}).
As for the SMC, we used the high-resolution H\,I data from the Australian Square Kilometre 
Array Pathfinder (ASKAP) obtained by \citet{diteodoroetal2019}. We compared the
NGC\,330's RV with that from the ASKAP velocity map (see their figure 1) for the cluster
position and found a very good agreement. 

\section{Conclusions}

We obtained Str\"omgren photometry of selected young MC clusters in order to
provide direct estimates of their metal contents, which are noticeably lacking
in the literature. The observations of 13 young MC clusters, namely: NGC\,330, 376,
1711, 1844, 1847, 1850, 1863, 1903, 1986, 2065, 2136, IC\,1611 and Lindsay\,35,
were performed with the SOI attached to the SOAR telescope during two observing runs in 
December 2008 and January 2009, respectively, as part of an observational programme aimed 
at studying the chemical evolution of these galaxies from their star clusters and
field star populations. 

In deriving the metallicities of measured yellow and red SGs we made use of an empirical
calibration recommended by \citet{gr1992}, based on the Str\"omgren metallicity sensitive 
index $m_{\rm 1}$. We paid particular attention in estimating the metallicity uncertainties,
which were calculated from propagation of all the involved errors added in quadrature, 
i.e., those coming from the obtained Str\"omgren photometry and those published from the 
employed metallcity calibration.

After a careful selection of yellow and red cluster SGs, on the basis of their
positions along the line-of-sight of the clusters, their locations in their respective 
cluster CMDs and relative placement in the cluster metallicty distribution functions,
we estimated mean cluster metallicities by applying a maximum likelihood approach.
The derived uncertainties are between 0.04 and 0.15 dex, with an average of 0.08 dex.
We found null intrinsic [Fe/H] spreads within the studied clusters with an upper
limit between 0.05 and 0.24 dex, with an average of 0.10 dex.

As far as we are aware, only NGC\,330 has previous metallicity estimates. Particularly, the most
recent [Fe/H] value obtained by \citet{gr1992} as well as those from high-dispersion
spectroscopy \citep{spiteetal1986} are in excellent agreement with that obtained
in this work. For the remaining studied clusters, the [Fe/H] values derived here are the first
metallicity estimates provided so far. In general, the resulting metal abundances agree
well with the known mean galaxy present-day metallicities, as expected since the youth of the
studied clusters. Nevertheless, there are some clusters whose derived mean [Fe/H] values
fall toward to edge of the present-day metallicity distribution function. We found that 
NGC\,300 and NGC\,1847 are at present the most metal-poor young clusters in the 
SMC and LMC, whereas NGC\,1711 one of the most  metal-rich in the LMC. 

When comparing the cluster metallicities with their present positions in the galaxies, 
we found evidence that support the outside-in formation scenario in both MCs. At the
same time, we found that interactions between the MCs and of the MCs with the Milky Way
could have caused that some clusters were scattered from their birthplaces. Indeed,
we show examples of LMC clusters with metal contents typical of the innermost galaxy regions
placed in the galaxy outer disc. Likewise, we found young clusters, at present located
in the inner regions of both MCs, formed out of gas that has remained unmixed since several
Gyr ago.

\section*{Acknowledgements}
We thank the referee for the thorough reading of the manuscript and
timely suggestions to improve it. 
This research has made use of the NASA/IPAC Extragalactic Database (NED), which is operated 
by the Jet Propulsion Laboratory, California Institute of Technology, under contract with 
the National Aeronautics and Space Administration.
This work has made use of data from the European Space Agency (ESA) mission
{\it Gaia} (\url{https://www.cosmos.esa.int/gaia}), processed by the {\it Gaia}
Data Processing and Analysis Consortium (DPAC,
\url{https://www.cosmos.esa.int/web/gaia/dpac/consortium}). Funding for the DPAC
has been provided by national institutions, in particular the institutions
participating in the {\it Gaia} Multilateral Agreement. We also thank
support from the IdP II 2015 0002 64 grant of the Polish Ministry of
Science and Higher Education.

\newpage



\bibliographystyle{mnras}

\input{paper.bbl}



\begin{figure*}
\caption{$V$ vs. $b-y$ CMDs for stars located within the cluster radius. Black 
filled circles represent stars with $(b-y)_o$ colours in the range 0.4 to 1.1 mag,
and which are brighter than the main sequence turnoff. Red filled circles
represent SGs used to estimate the cluster metallicity; the right-hand panels show their positions with error bars
in the $(m_{\rm 1})_o$ vs. $(b-y)_o$ plane, with the \citet{gr1992}'s iso-abundance lines 
(eq. (1)) superimposed (see details in Section 3).}
     \includegraphics[width=\columnwidth]{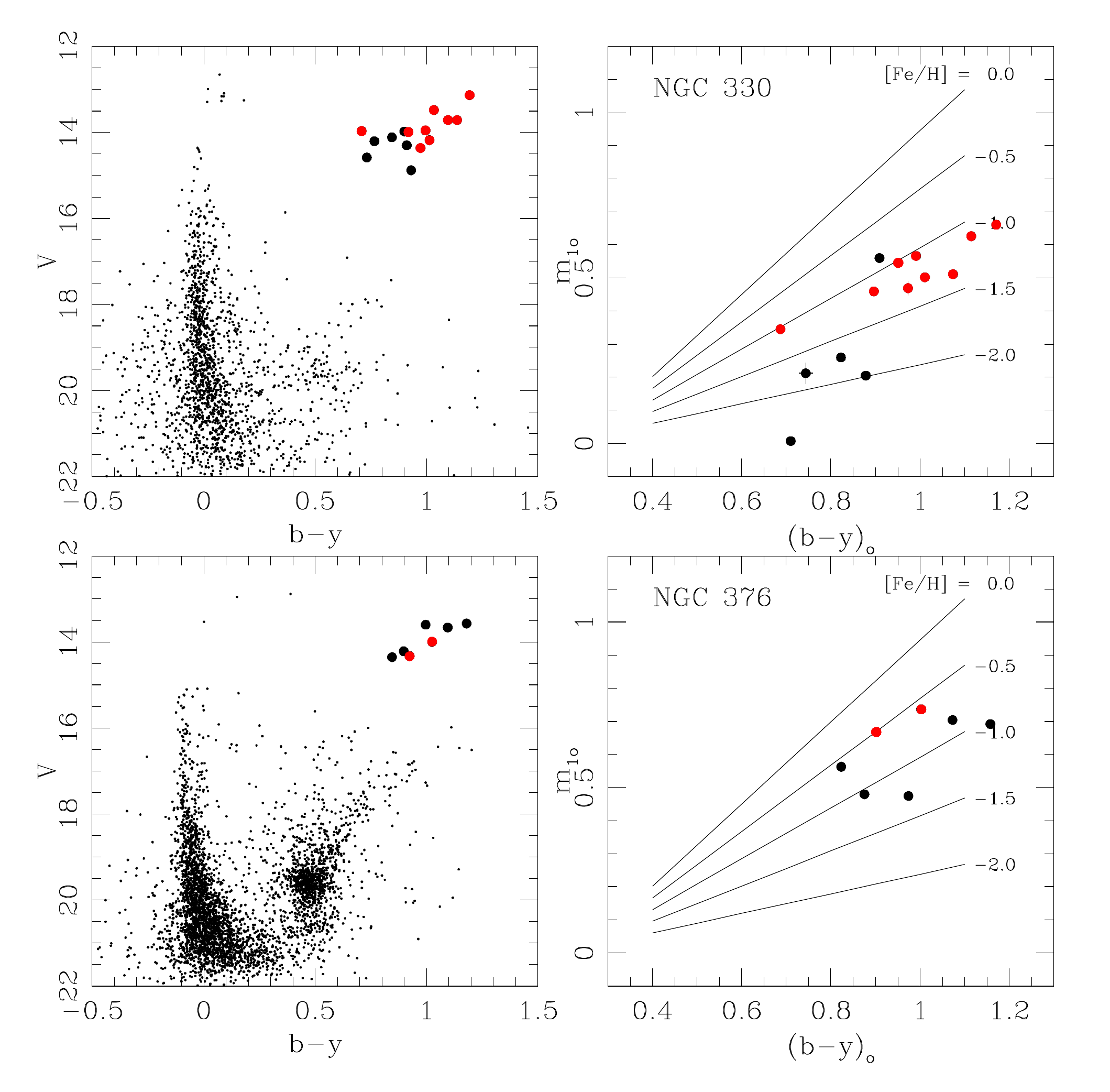}
     \includegraphics[width=\columnwidth]{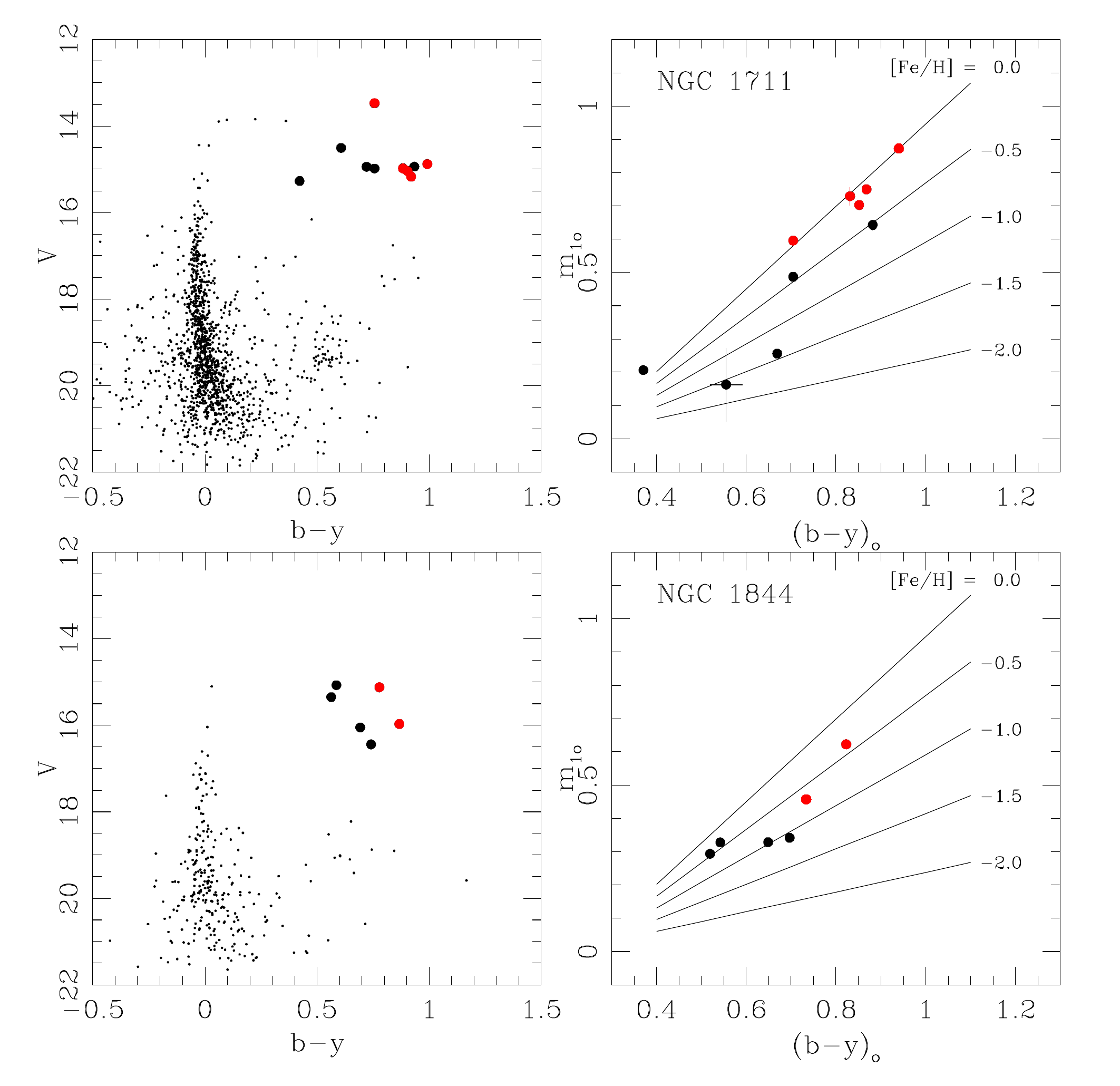}
     \includegraphics[width=\columnwidth]{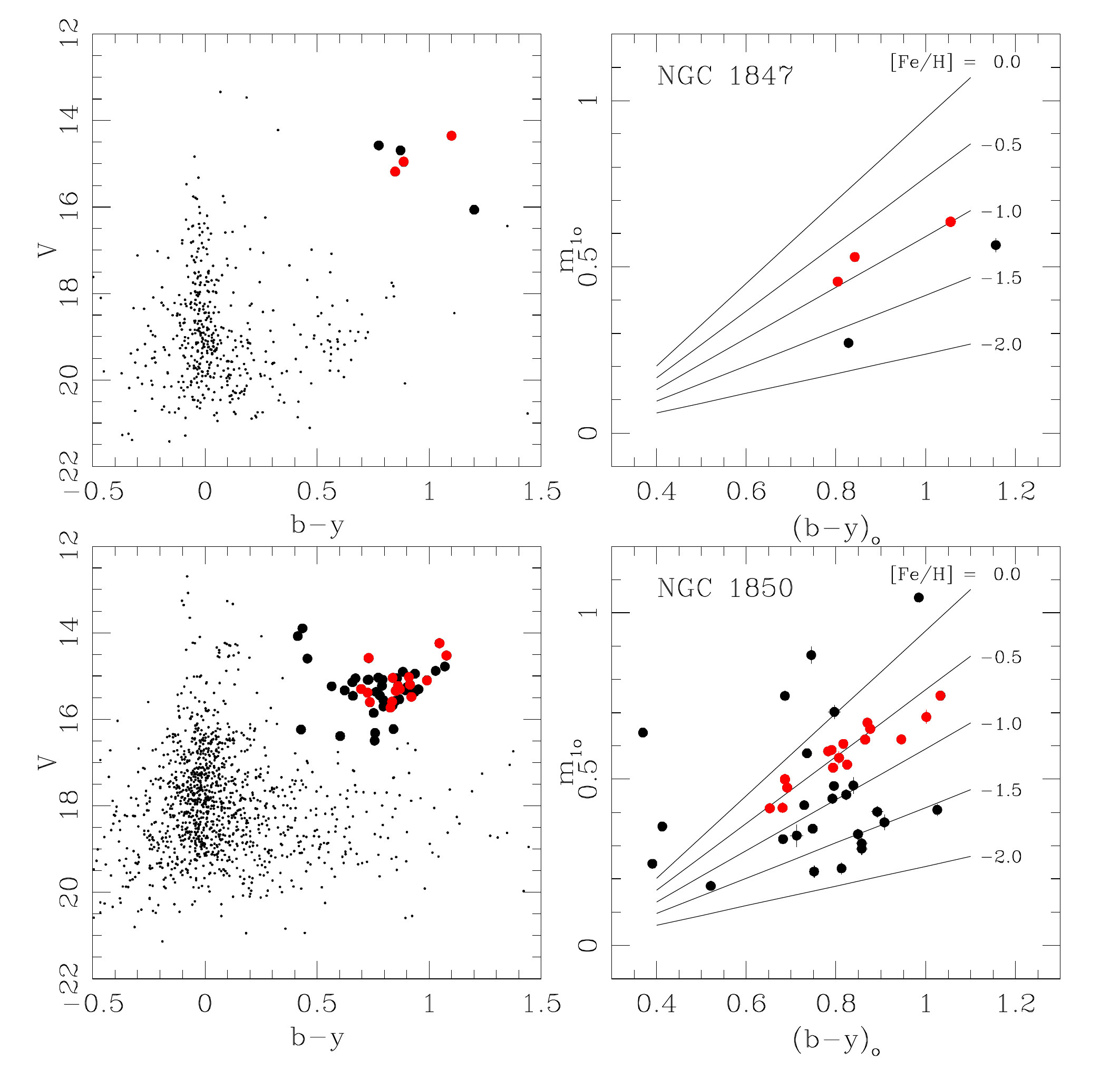}
     \includegraphics[width=\columnwidth]{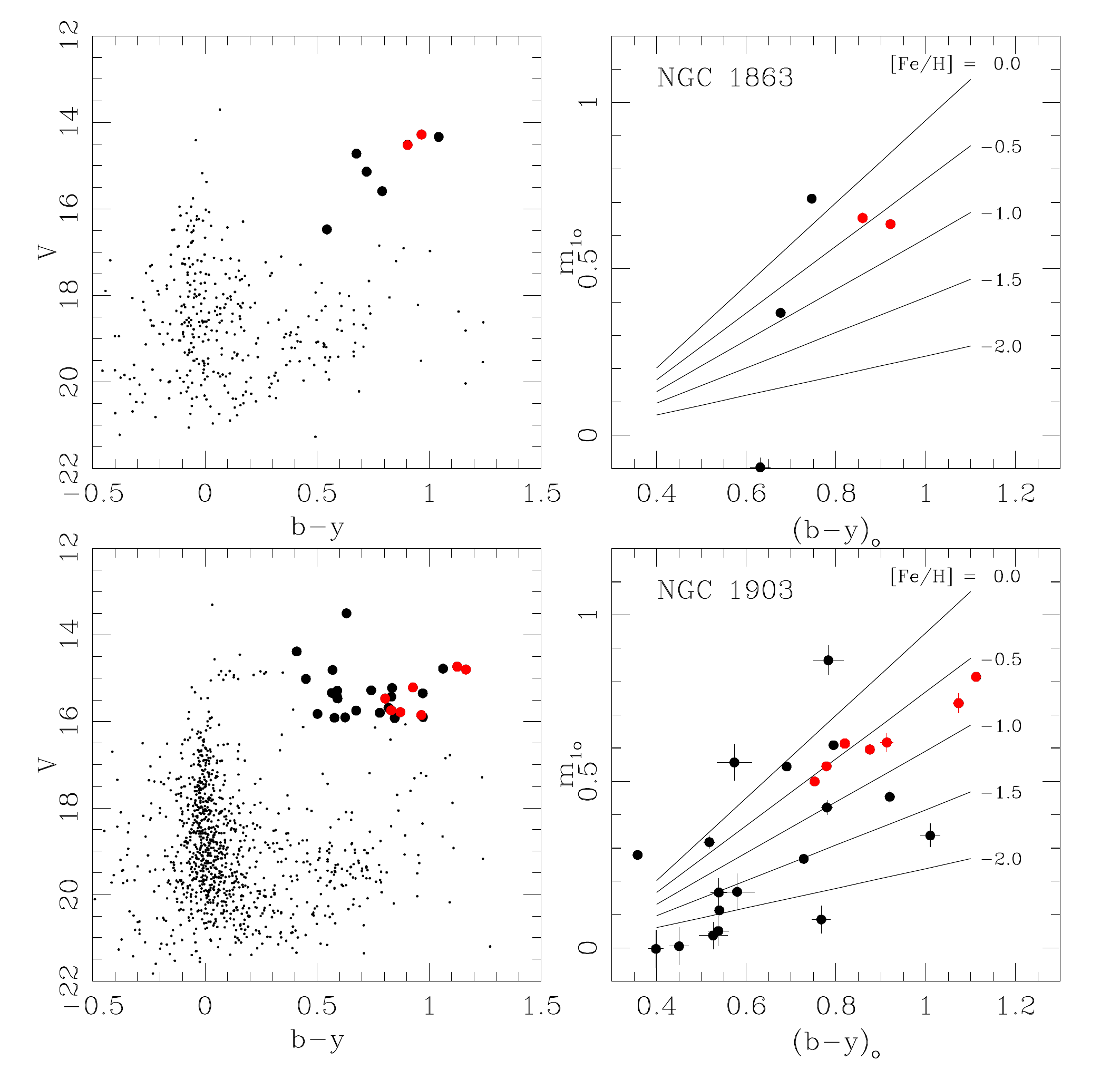}
 \label{fig:fig1}
\end{figure*}

\setcounter{figure}{0}
\begin{figure*}
\caption{continued.}
     \includegraphics[width=\columnwidth]{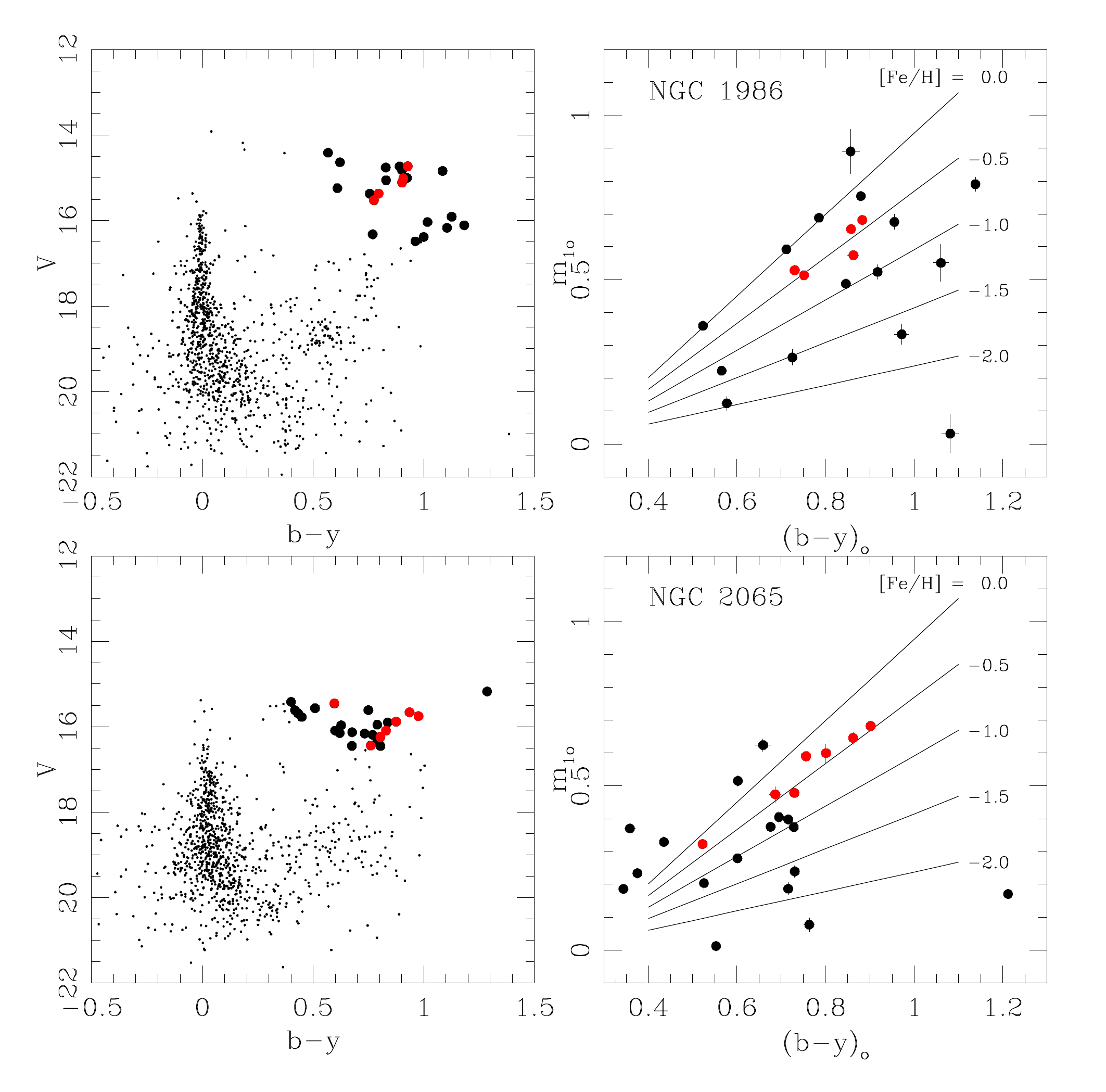}
     \includegraphics[width=\columnwidth]{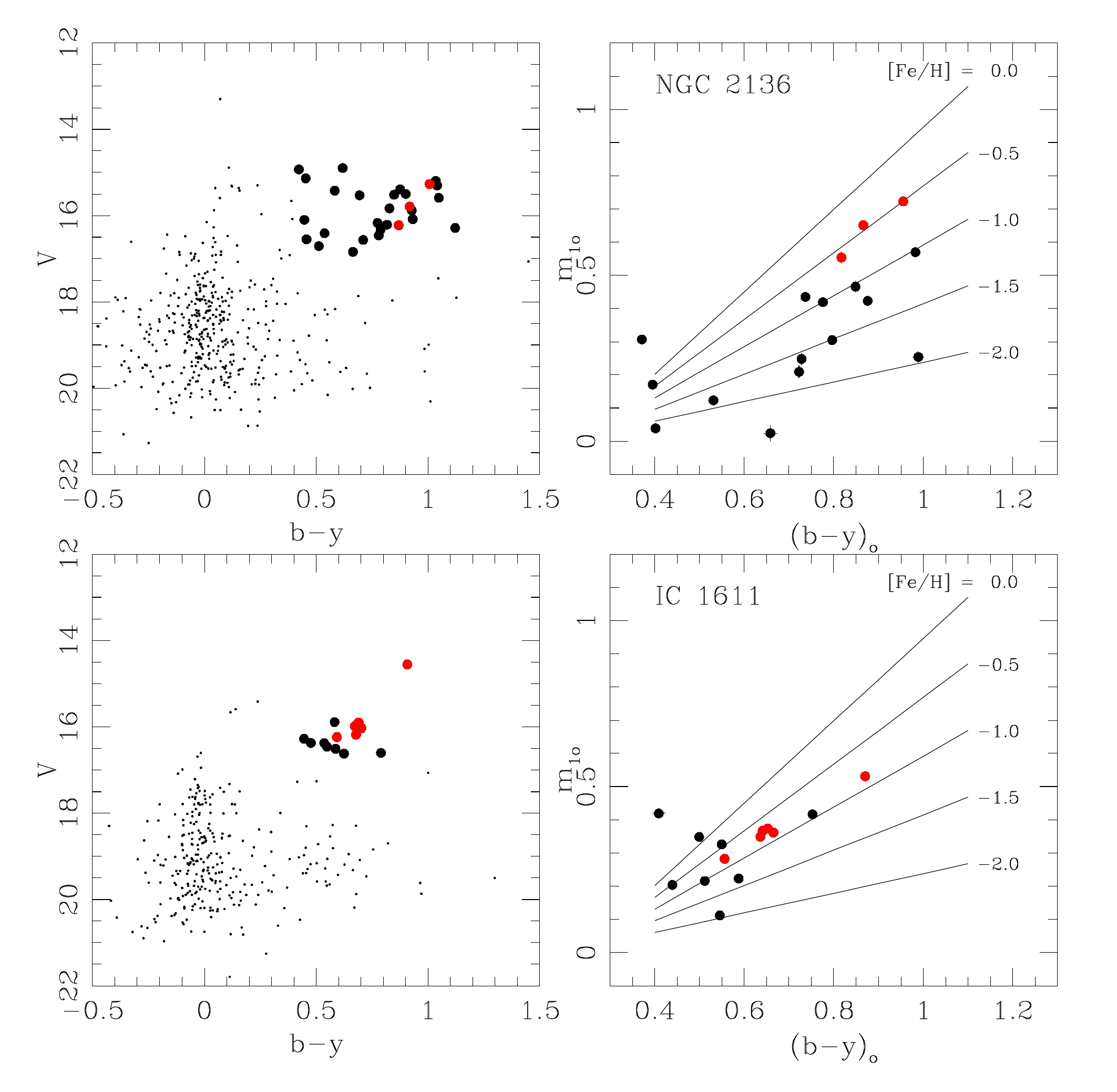}
     \includegraphics[width=\columnwidth]{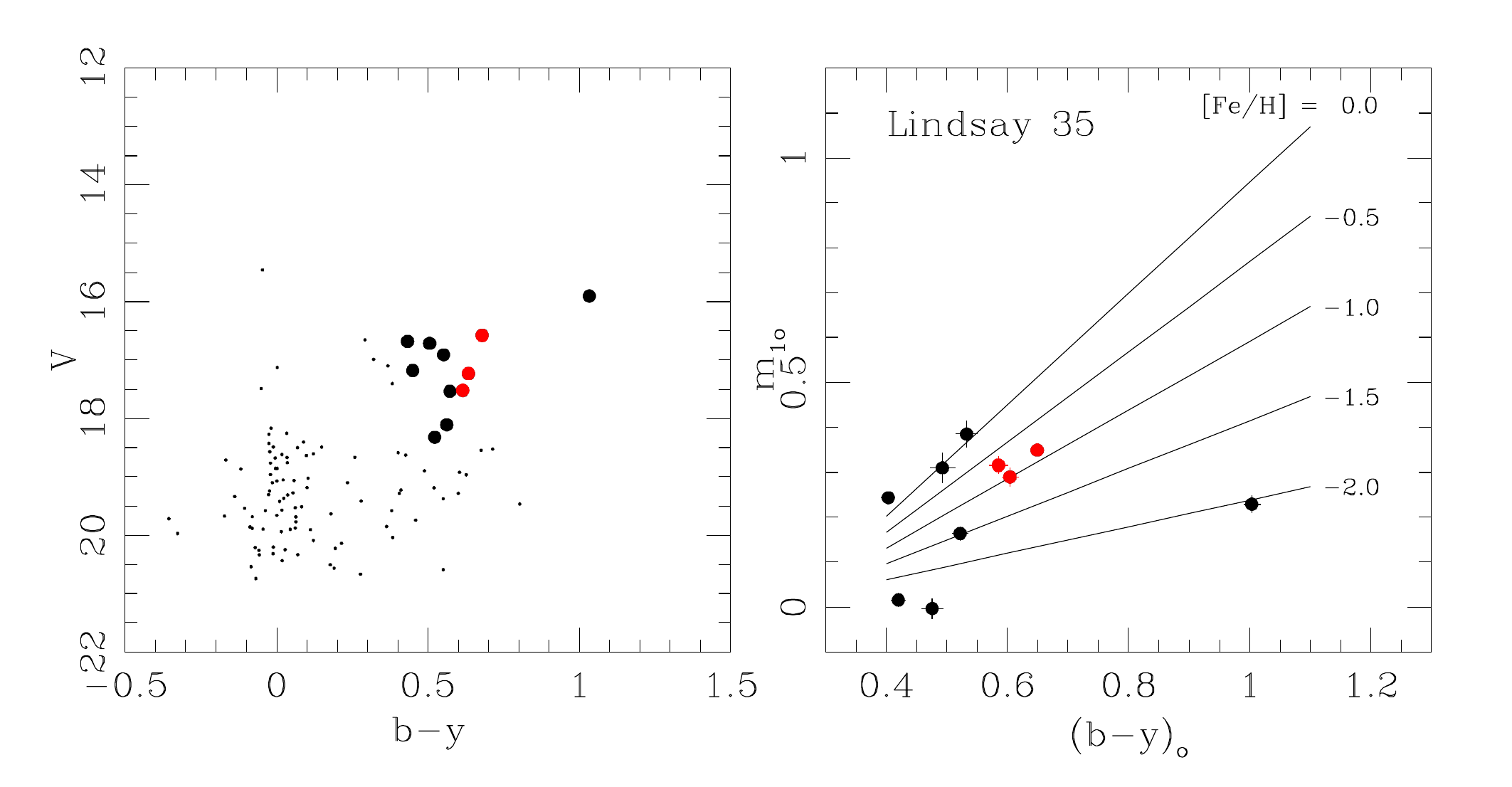}
\end{figure*}

\begin{figure*}
\caption{Spatial distribution of MC clusters. The \citet{hz09}'s LMC regions and 
the \citet{petal07d}'s SMC elliptical framework are superimposed. The contour of the LMC bar is
delineated with a light-blue line, while the SMC ellipses have semi-major axes of
1, 2 and 4 degrees, respectively. The clusters in the
\citet{betal08}'s catalogue are drawn with black points. Light green, red and orange
big circles represent NGC\,330, NGC\,1711 and NGC\,1847, respectively.}
     \includegraphics[width=\textwidth]{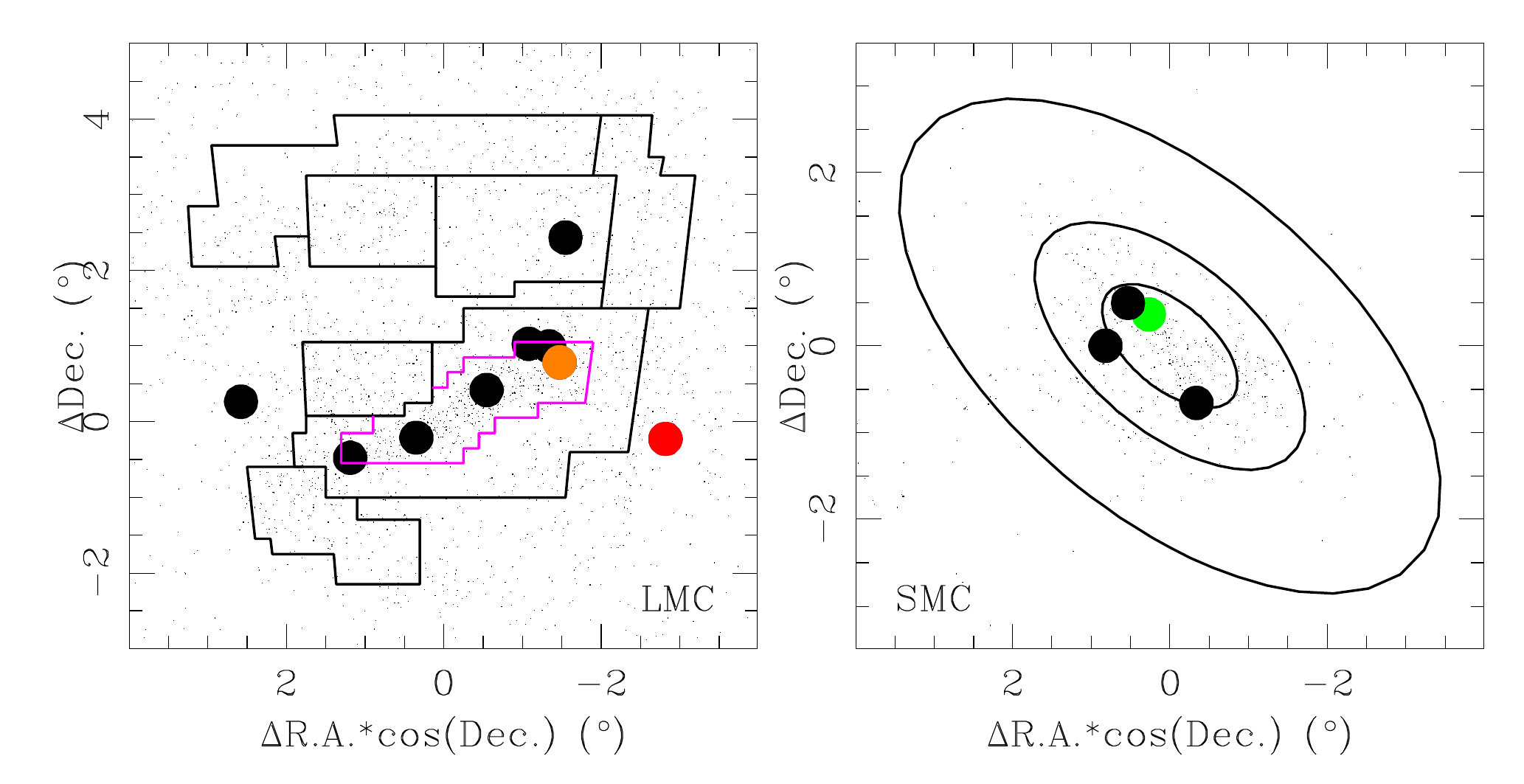}
 \label{fig:fig2}
\end{figure*}

\begin{table}
\caption{Log of observations.}
\label{tab:table1}
\begin{tabular}{@{}lccccccc}\hline
Cluster & Date & \multicolumn{3}{c}{Exposure time (sec)} & \multicolumn{3}{c}{Airmass} \\
        &      &   $v$   &  $b$  &  $y$        &  $v$  & $b$  &  $y$ \\\hline
NGC\,330    & 17 Dec. 2008 & 400 & 200 & 120 & 1.50 &1.49 &1.49\\
            & 19 Dec. 2008 & 400 & 160 & 100 & 1.48 &1.48 &1.47\\
NGC\,376    & 18 Dec. 2008 & 500 & 300 & 180 & 1.53 &1.52 &1.52\\
            & 19 Dec. 2008 & 350 & 140 &  90 & 1.53 &1.52 &1.52\\
NGC\,1711   & 16 Jan. 2009 & 350 & 180 & 100 & 1.37 &1.38 &1.38\\
NGC\,1844   & 17 Jan. 2009 & 400 & 140 & 100 & 1.27 &1.27 &1.27\\
NGC\,1847   & 17 Jan. 2009 & 350 & 160 & 100 & 1.29 &1.30 &1.30\\
NGC\,1850   & 18 Jan. 2009 & 400 & 160 & 100 & 1.34 &1.34 &1.34\\
NGC\,1863   & 18 Jan. 2009 & 400 & 180 & 100 & 1.32 &1.32 &1.32\\
NGC\,1903   & 17 Dec. 2008 & 300 & 100 &  60 & 1.36 &1.36 &1.36\\
NGC\,1986   & 18 Jan. 2009 & 350 & 180 & 100 & 1.53 &1.52 &1.51\\
NGC\,2065   & 18 Jan. 2009 & 400 & 180 &  90 & 2.07 &2.05 &2.04\\
NGC\,2136   & 17 Jan. 2009 & 450 & 180 & 110 & 1.99 &1.97 &1.96\\
IC\,1611    & 18 Dec. 2008 & 500 & 200 & 120 & 1.57 &1.56 &1.56\\
Lindsay\,35 & 18 Dec. 2008 & 500 & 200 & 120 & 1.83 &1.81 &1.81\\\hline
\end{tabular}
\end{table}

\begin{table*}
\caption{Transformation coefficients to the Str\"romgren photometric system.}
\label{tab:table2}
\begin{tabular}{@{}ccccccc}\hline
Date   &  Filter   &  coef$_1$  &  coef$_2$ &  coef$_3$ &  coef$_4$ & rms \\\hline
17 Dec. 2008 & $y$ & 0.946      &  0.118      &  -0.008      &  &  0.010 \\
             &     & $\pm$0.015 &  $\pm$0.009 &  $\pm$0.015  &  &        \\
             & $b$ & 0.959      &  0.163      &  0.942       &  &  0.002 \\
             &     & $\pm$0.003 &  $\pm$0.002 & $\pm$0.003   &  &         \\
             & $v$ & 1.137      &  0.301      &  2.008       &  1.028   &  0.017\\
             &     & $\pm$0.027 &  $\pm$0.016 &  $\pm$0.058  & $\pm$0.068  &   \\
18 Dec. 2008 & $y$ & 0.932      &  0.122      & -0.005       &  &  0.010  \\
             &     & $\pm$0.015 &  $\pm$0.009 &  $\pm$0.016  &  &          \\
            & $b$ & 0.942      &  0.177      &  0.946       &  & 0.008  \\
            &     & $\pm$0.014 &  $\pm$0.009 &  $\pm$0.014  &  &         \\
             & $v$ & 1.122      &  0.295      &  1.995       & 1.026  &  0.002 \\
             &     & $\pm$0.007 &  $\pm$0.005 &  $\pm$0.048  & $\pm$0.061  &    \\
19 Dec. 2008 & $y$ & 0.939      &  0.107      &  0.018       &  & 0.016 \\
             &     & $\pm$0.019 &  $\pm$0.010 &  $\pm$0.015  &  &        \\
            & $b$ & 0.916      &  0.169      &  0.999       &  &  0.010  \\
            &     & $\pm$0.013 &  $\pm$0.007 &  $\pm$0.011  &  &          \\
             & $v$ & 1.096      &  0.286      &  2.004       &  1.117  &  0.010 \\
            &     & $\pm$0.015 &  $\pm$0.009 &  $\pm$0.030  &  $\pm$0.038  &     \\
16 Jan. 2009 & $y$ & 1.005      &  0.120      & -0.046       &  &    0.007 \\
             &     & $\pm$0.004 &  $\pm$0.010 &  $\pm$0.011  &  &         \\
            & $b$ & 1.014      &   0.170     &  0.939       &  &     0.011\\
            &     & $\pm$0.007 &  $\pm$0.003 &  $\pm$0.018  &  &         \\
            & $v$ & 1.196      &  0.290      &  2.034       & 0.914  &   0.007 \\
            &     & $\pm$0.005 &  $\pm$0.010 & $\pm$0.032        & $\pm$0.028 &   \\
17 Jan. 2008 & $y$ & 0.940      &  0.155      &  0.012       &  &     0.017  \\
             &     & $\pm$0.019 &  $\pm$0.023 &  $\pm$0.090       &  &       \\
            & $b$ & 0.957      &  0.201      &  0.931       &  &   0.013  \\
             &     & $\pm$0.015 &  $\pm$0.014 &  $\pm$0.058  &  &      \\
             & $v$ & 1.194      &  0.295      &  2.025       &  0.950 & 0.010 \\
             &     & $\pm$0.010 &  $\pm$0.007 &  $\pm$0.049  & $\pm$0.058  &    \\
18 Jan. 2008 & $y$ & 1.003      &  0.132      & -0.035       &  &  0.013 \\
             &     & $\pm$0.013 &  $\pm$0.007 &  $\pm$0.023  &  &     \\
             & $b$ & 1.013      &  0.184      &  0.916       &  &   0.008\\
             &     & $\pm$0.008 &  $\pm$0.004 &  $\pm$0.014  &  &   \\
             & $v$ & 1.194      &  0.300      &  2.018       & 0.987  &   0.012 \\
             &     & $\pm$0.032 &  $\pm$0.016 &  $\pm$0.097  & $\pm$0.092  &   \\\hline
\end{tabular}
\end{table*}

\begin{table*}
\caption{Astrophysical properties of MC clusters.}
\label{tab:table3}
\begin{tabular}{@{}lcccccc}\hline
Cluster &  \multicolumn{2}{c}{$E(B-V)$ (mag)} &   Age (Myr)   &  [Fe/H] (dex) & Ref.  & [Fe/H] (dex) \\
        & H11  & NED  &         &           &      &       \\ 
\hline
\multicolumn{7}{c}{SMC} \\\hline
NGC\,330    &--- &  0.03 &   40 & -0.90 & 1   &   -1.15$\pm$0.06 \\
NGC\,376    & 0.03&  0.03 &   28 & -0.60 & 2  &   -0.55$\pm$0.09\\
IC\,1611    & 0.05&  0.03 &  100 & -0.70 & 6   &   -0.80$\pm$0.09 \\
Lindsay\,35 & 0.04&  0.03 &  220 & -0.70 &11   &   -0.85$\pm$0.15 \\ \hline
\multicolumn{7}{c}{LMC} \\\hline
NGC\,1711   & 0.07&  0.06 &   50 & -0.57  & 12   &   -0.06$\pm$0.05 \\
NGC\,1844   & 0.04&  0.06 &  150 & -0.20 & 4   &   -0.50$\pm$0.11 \\
NGC\,1847   & 0.05&  0.06 &   50 & -0.40 & 3   &   -0.91$\pm$0.09 \\
NGC\,1850   & 0.06&  0.06 &   80 & -0.40 & 5   &   -0.53$\pm$0.04 \\
NGC\,1863   & 0.05&  0.06 &   40 & -0.40 & 8   &   -0.53$\pm$0.09 \\
NGC\,1903   & 0.07&  0.06 &  100 & -0.40 & 9   &   -0.60$\pm$0.05 \\
NGC\,1986   & 0.05&  0.06 &   70 & ---  &10   &   -0.46$\pm$0.06 \\
NGC\,2065   & 0.10&  0.06 &  100 & ---  & 7   &   -0.40$\pm$0.06 \\
NGC\,2136   & 0.07&  0.06 &  124 & -0.50 & 3   &   -0.51$\pm$0.08 \\
\hline
\end{tabular}

\noindent Ref.: (1) \citet{miloneetal2018}; (2) \citet{sabbietal2011}; (3) 
\citet{niederhoferetal15a}; (4) \citet{miloneetal13}; (5) \citet{bastianetal2017}; 
(6) \citet{petal07d}; (7) \citet{asadetal2016}; (8) \citet{petal03}; (9) \citet{petal15b};
(10) \citet{ef1985}; (11) \citet{pietal08}; (12) \citet{dirschetal2000}.
\end{table*}





\bsp	
\label{lastpage}
\end{document}